\documentclass[12pt]{iopart}

\usepackage{graphicx}% Include figure files
\usepackage{subfig}
\usepackage{tabularx}
\usepackage{dcolumn}% Align table columns on decimal point
\usepackage{bm}% bold math
\usepackage{braket}
\usepackage[justification=raggedright]{caption}
\begin{document}
	
	\title{Resistive method for measuring the disintegration speed of Prince Rupert's drops}% Force line breaks with \\
	
	\author{Daria Gusenkova, Mark Bochkov, Evgenii Glushkov, Julia Zotova and S. N. Zhabin}
	
	\address{Moscow Institute of Physics and Technology}
	
	\ead{eugene.glushkov@phystech.edu}

	\date{\today}% It is always \today, today,
	%  but any date may be explicitly specified
	
	\begin{abstract}
		We have successfully applied the resistance grid technique to measure the disintegration speed in special type of glass objects, widely known as Prince Rupert's drops. We use a digital oscilloscope and a simple electrical circuit, glued to the surface of the drops, to detect the voltage changes, corresponding to the breaks in the specific parts of the drops. The results obtained using this method are in good qualitative and quantitative agreement with theoretical predictions and previously published data. Moreover, the proposed experimental setup doesn't include any expensive equipment (such as a high-speed camera) and can therefore be widely used in high schools and universities.
		
	\end{abstract}
	
	
	
	\maketitle
	
	%\tableofcontents
	
	\section{\label{sec:intro}Introduction}
	Prince Rupert's drops are widely known tear-shaped glass objects with a thin tail obtained by dropping hot molten glass into water. During the process of their formation the surface of the molten glass is quickly cooled, while the inner portion of the drop remains significantly hotter.  After complete cooling this leads to large compressive stresses on the surface, while the core of the drop is in the state of tensile stress. 
	
	The earliest study of Prince Rupert's drops was performed by Robert Hooke after they were introduced to the Royal Society of London in 1660 by King Charles II. These glass objects were named after his nephew, Prince Rupert of Bavaria, who has brought these droplets of molten glass from Germany to England and shown them to the King. Later the detailed illustration of Prince Rupert's drop appeared in Hooke's \mbox{\textit{Micrographia}}~\cite{hooke1665}, where he described the process of its formation and cooling. The further history of these peculiar objects can be found in a review by Brodsley et al. \cite{brodsley1986}, while experimental and analytical characterization of temperatures, residual stresses and densities of Rupert's drops for various types of glass and at different stages of formation can be found in the paper by Johnson and  Chandrasekar \cite{Johnson1992}. 
	
	One special feature, that Prince Rupert's drops possess, is the ability to withstand large mechanical pressure, applied to their head, without any deformation and turn into glass powder after smallest crack at the tail of the droplet. After the initial crack is formed, the destruction process moves from the tail to the head of the drop with high speed (in the order of kilometers per second). 
	
	%TODO write about other methods of measuring the speed 
	First precise measurements of this disintegration speed were performed by Chandrasekar and Chaudri \cite{Chandrasekar1994} using a high-speed camera, which was able to shoot up to half a million frames per second. This is an expensive piece of equipment, which is rarely available for regular students at most universities, especially in the developing countries. That is why we decided to apply another method, known as resistance grid technique \cite{Anthony1970} to measure the disintegration velocity of Prince Rupert's drops and investigate how it depends on the environmental conditions during the drop formation and properties of the glass used. This experiment was initially designed and performed for the International Physicists' Tournament - a team competition for students \cite{vanovskiy2014}, which was held in Warsaw in April, 2015.

	\section{\label{sec:theory}Theoretical overview}
	
	In this brief overview we focus on the physical picture behind the peculiar cracking behavior of Prince Rupert's drops. As described above, the ability of Prince Rupert's drops to withstand large mechanical pressure and quickly disintegrate after breaking their tail is due to the process of their formation, when a liquid glass drop flies through air and comes in contact with water. After cooling quickly, the outside of the drop becomes compressed, while the inner part is under tension. This leads to a large amount of elastic energy, stored in the drop, that is released when the tail is broken, which leads to fast disintegration of the whole drop. 
	
	There were several attempts to theoretically estimate the disintegration speed of Prince Rupert's drops, but the exact mechanism behind the cracking process still remains the topic of ongoing research. The foundation of the crack's dynamics in brittle materials was set by Griffith \cite{griffith1921} and Mott \cite{mott1948}, followed by the work of Yoffe \cite{yoffe1951}, and the estimation of the crack's limiting velocity was given by Roberts and Wells \cite{roberts1954} in the case of static stress ($0.38 c_0$, where $c_0=\sqrt{E/\rho}$ is the longitudinal sound velocity) and by Steverding and Lehnigk \cite{steverding1970} in the case of pulsed stress ($0.52 c_R$, where $c_R$ is the velocity of the Rayleigh surface waves). All this work gave rise to the field of fracture mechanics, which was recently reviewed by Bouchbinder et al. \cite{bouchbinder2010}.
	
	To address the question of the complicated structure of the crack, which looks similar to fractals, another approach was used, called fractal fracture mechanics \cite{mosolov1991, cherepanov1995}. The estimation of the propagation speeed of the crack in the framework of fractal fracture mechanics was given by Yavari and Khezrzadeh \cite{Yavari2010} in the range $\left[0.318 c_0,0.321 c_0\right]$.
	
	Prince Rupert's drops can be also viewed as a specific example of a broader phenomena, know as failure waves or self-sustaining fracture waves \cite{galin1966, cherepanov2009, bless2010}, that occurs in glass and other brittle materials.
	
	However, unlike ordinary tempered glass, which breaks into small cubical fragments, Prince Rupert's drops explode upon rupture and the precise mechanism of this disintegration still hasn't been explained. The high-speed photographic studies \cite{Chandrasekar1994,Chaudhri1998} suggested crack bifurcation as the main mechanics and the experimental statistical analysis of the disintegrated drops, done by Silverman et al. \cite{silverman2012}, tested various theoretical fragmentation models that predict the form of the particle, mass densities and the fractal dimension of the set of fragments. An important contribution to finding the disintegration mechanism of Prince Rupert's drops comes from computer simulations of such structures \cite{herrmann1989,herrmann1991}. The next step to answer this puzzle would be to investigate drops, exploded within some sort of confining matrix to permit statistical analysis of fragments drawn separately from areas of tension and compression.
	
	\section{\label{sec:exper}Experiment}
	
	\subsection{\label{subsec:fabrication}Fabrication of samples}
	To produce the drops we heated glass sticks with a gas burner. The temperature of the flame was crucial at this point – it had to be greater than the melting temperature and less than boiling temperature of the glass used. If the temperature of the flame is too high the process of boiling creates local irregularities in the structure of the drop, which causes its immediate destruction upon contact with cooler medium.
	
	%which temperature and burner did you use?
	
	\begin{figure}[ht]
		\centering
		\subfloat[]{
			\label{fig:fig1-a}
			\includegraphics[width=.4\columnwidth]{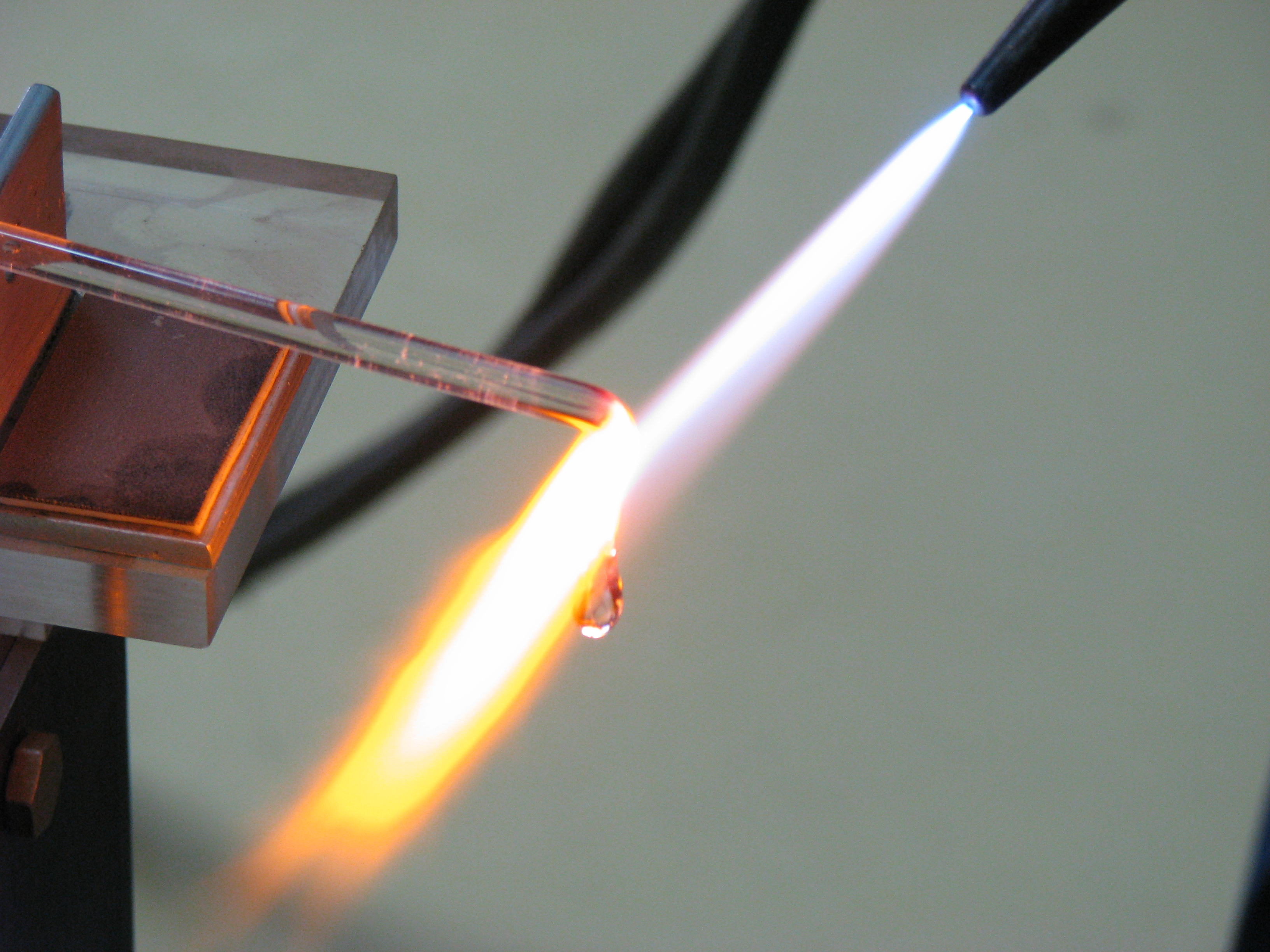}}
		\subfloat[]{
			\label{fig:fig1-b}
			\includegraphics[width=.4\columnwidth]{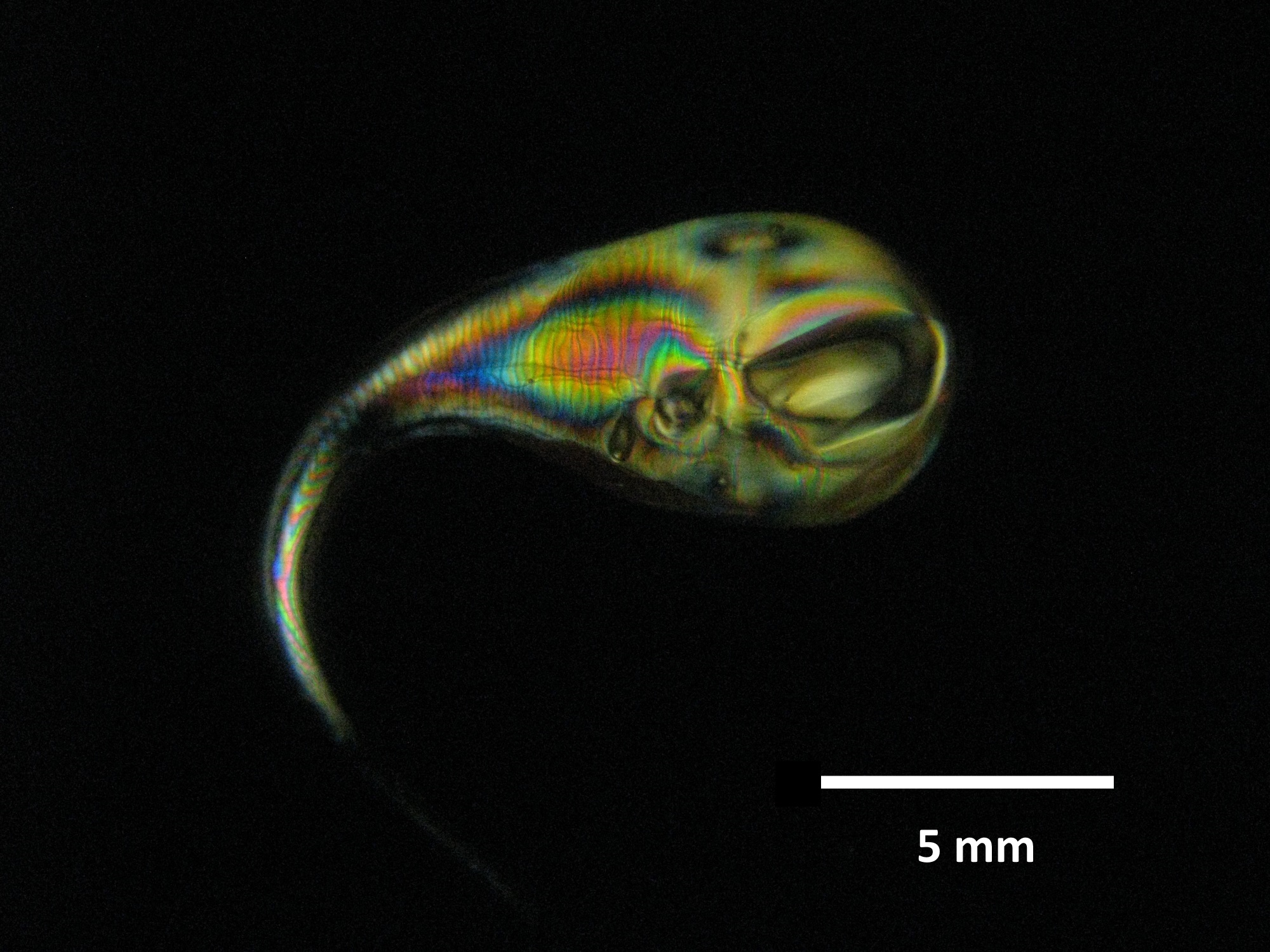}}
		\caption{Fabrication of the Prince Rupert's drop by melting a glass stick with a gas burner (a) and the residual stresses in the drop after cooling (b).}
		\label{fig:fabrication}
	\end{figure}
	
	After melting the drop falls into a tank, filled with liquid, and cools in it.	We used four different types of glass (soda-lime, borosilicate, blue- and red-colored) to produce the drops and both water (warm and cold) and liquid nitrogen to cool the drops. The colored drops contained some inclusions of heavier elements: cobalt for blue glass and selenium for red glass. The stress distribution in the created Prince Rupert's drops was obtained using crossed polarizers and is shown in the figure~\ref{fig:fig1-b}.
	
	\subsection{\label{subsec:setup}Experimental setup}
	
\begin{figure}[ht]
		\centering
		\subfloat[]{
			\label{fig:fig2-a}
			\includegraphics[width=.4\columnwidth]{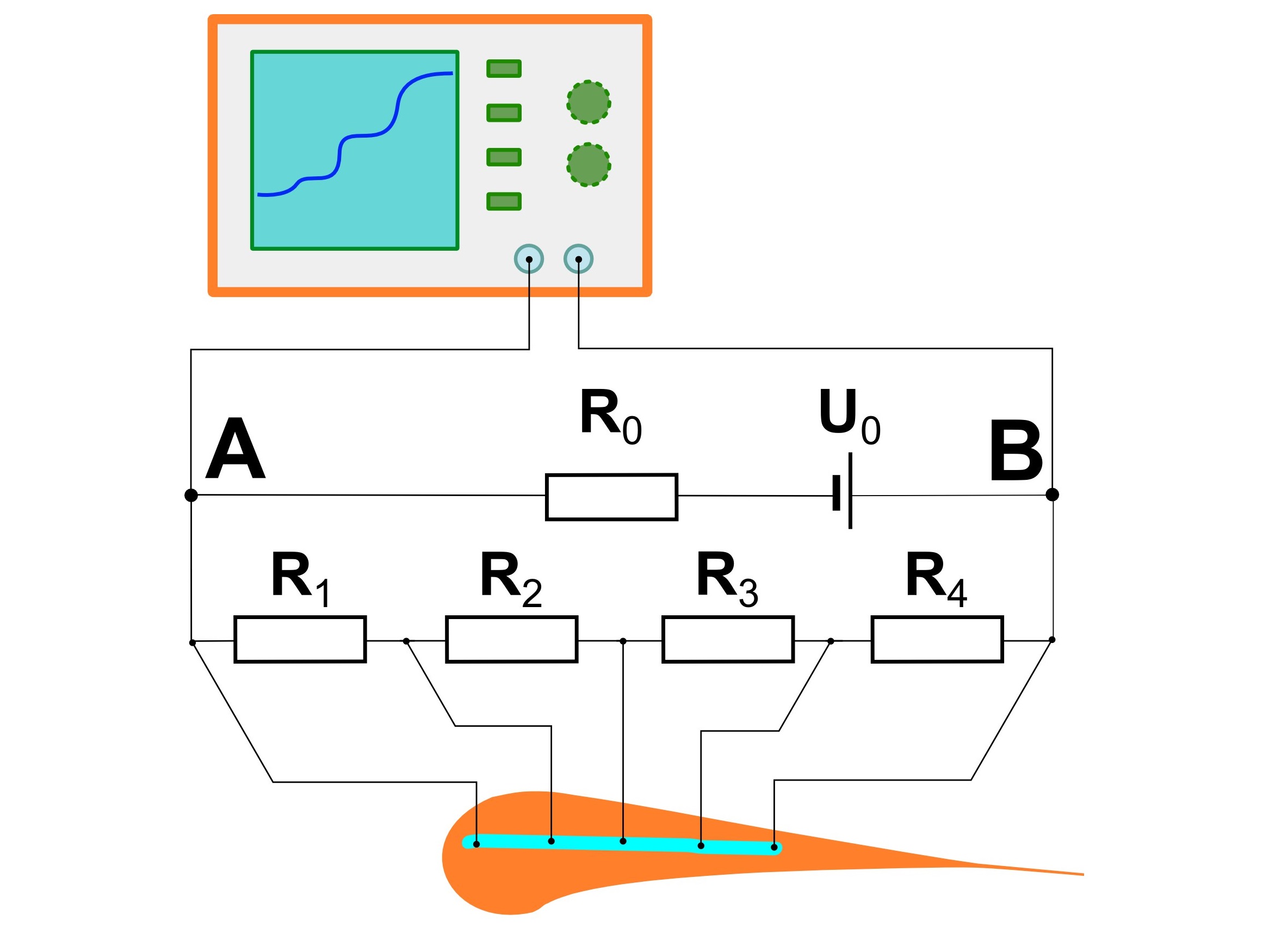}}
		\subfloat[]{
			\label{fig:fig2-b}
			\includegraphics[width=.4\columnwidth]{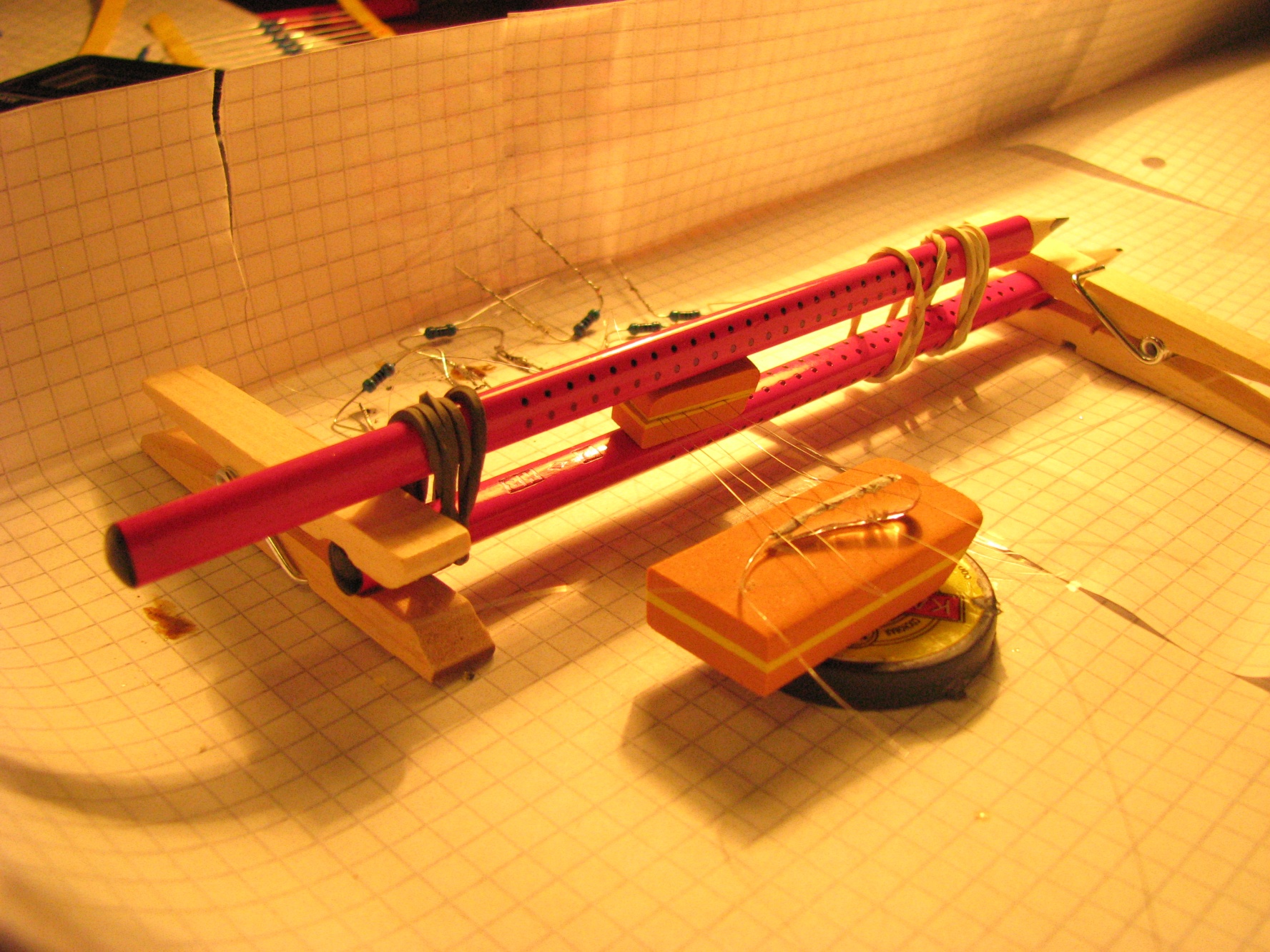}}
		\caption{Proposed experimental setup for measuring the disintegration velocity of Prince Rupert's drops (a) and its first hands-on implementation (b).}
		\label{fig:fig2}
	\end{figure}

	The main idea behind the method used to measure the disintegration speed of Prince Rupert's drops is to convert breaking of glass into an electrical signal that could be easily observed with an average oscilloscope. This can be achieved by a simple electric circuit with resistors that could be turned "on" or "off" by the propagating fracture wave. The proposed setup is schematically shown in figure~\ref{fig:fig2-a}.
	
	It consists of a voltage source driving current through a chain of resistors connected in series. The first resistor limits overall current in the circuit. All other resistors are connected to a common conductive bus that is formed on the surface of a measured sample. A digital oscilloscope measures the overall voltage drop on control resistors $R_1 ... R_4$. Nominal resistance for all resistors used in the experiment was chosen to be 1~KOhm.
	
	In the initial state all electrical current flows through the bus on the drop's surface ignoring control resistors (provided their resistance is high in comparison with the resistance of the bus). When the drop is disintegrating the fracture wave moves from its tail and gradually breaks conductive bus, which sequentially adds control resistors into the circuit. That's why electric current starts to flow through resistors and it causes abrupt increase of the control voltage drop.
	
	The assembled experimental setup is shown in figure~\ref{fig:fig2-b}. It consists of an electrical circuit with wires attached to control resistors hanging freely in the air above the mount point of a studied drop. Those wires are attached to the drop's surface using a conductive glue. Drops are fixed to the base with cyan-acrylic glue in order to maintain the same position during preliminary preparations. The conductive bus is also formed with the same conductive glue on the surface of the drop, connecting all wires. %which voltage source was used?
	
	After the setup was ready for taking measurements, we initiated the disintegration process by breaking the free-hanging tail of the drop with a wire-cutter. An oscilloscope (RIGOL DS1102E) with time span of 200 ns and temporal resolution of 4 ns was used to record changes of electrical flow through control resistors. Knowing the distance between control wires and time taken by each step it is possible to estimate the propagation speed of the fracture wave.
	
	% Describe, how it was fitted!
	\begin{figure}[h]
		\centering
		\includegraphics[width=0.8\columnwidth]{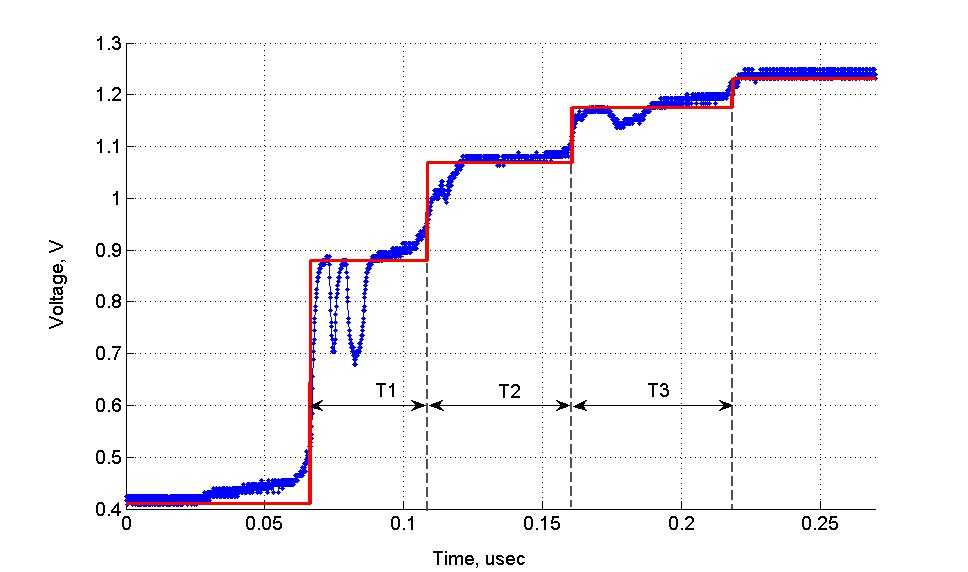}
		\caption{Processed signal from the digital oscilloscope, fitted using \ref{eq:R1}. }
		\label{fig:fig3}
	\end{figure}

	\subsection{\label{subsec:data}Measured data}
	
	With the digital oscilloscope we collected several sets of data, the typical look of which is presented in figure~\ref{fig:fig3}. One can see distinct voltage steps, each of which corresponds to the next section of wires being torn away. Knowing the lengths of conductive sections and time of their destruction we can estimate Prince Rupert's drop disintegration speed. 
	
	To fit the obtained individual data traces (red line in figure~\ref{fig:fig3}) we used the following expression, taking into account non-zero bus resistance:
	
	\begin{equation}\label{eq:R1}
	U_{sig} = \frac {U_0 R_{sum}}{R_0+R_{sum}},
	\end{equation}
	
	\noindent where $R_{sum}$ is the equivalent resistance of the control resistors in parallel to the finite resistance of the conductive bus, which varies for different samples. For the data set, shown in figure~\ref{fig:fig3}, the values of additional bus resistances are $r_0=0.3~k\Omega$, $r_1 = r_2 = r_3 = 0.05~k\Omega$, where $r_i$ is the resistance of the corresponding bus segment.
	
\begin{figure}[ht]
		\centering
		\subfloat[]{
			\label{fig:fig4-a}
			\includegraphics[width=0.33\textwidth]{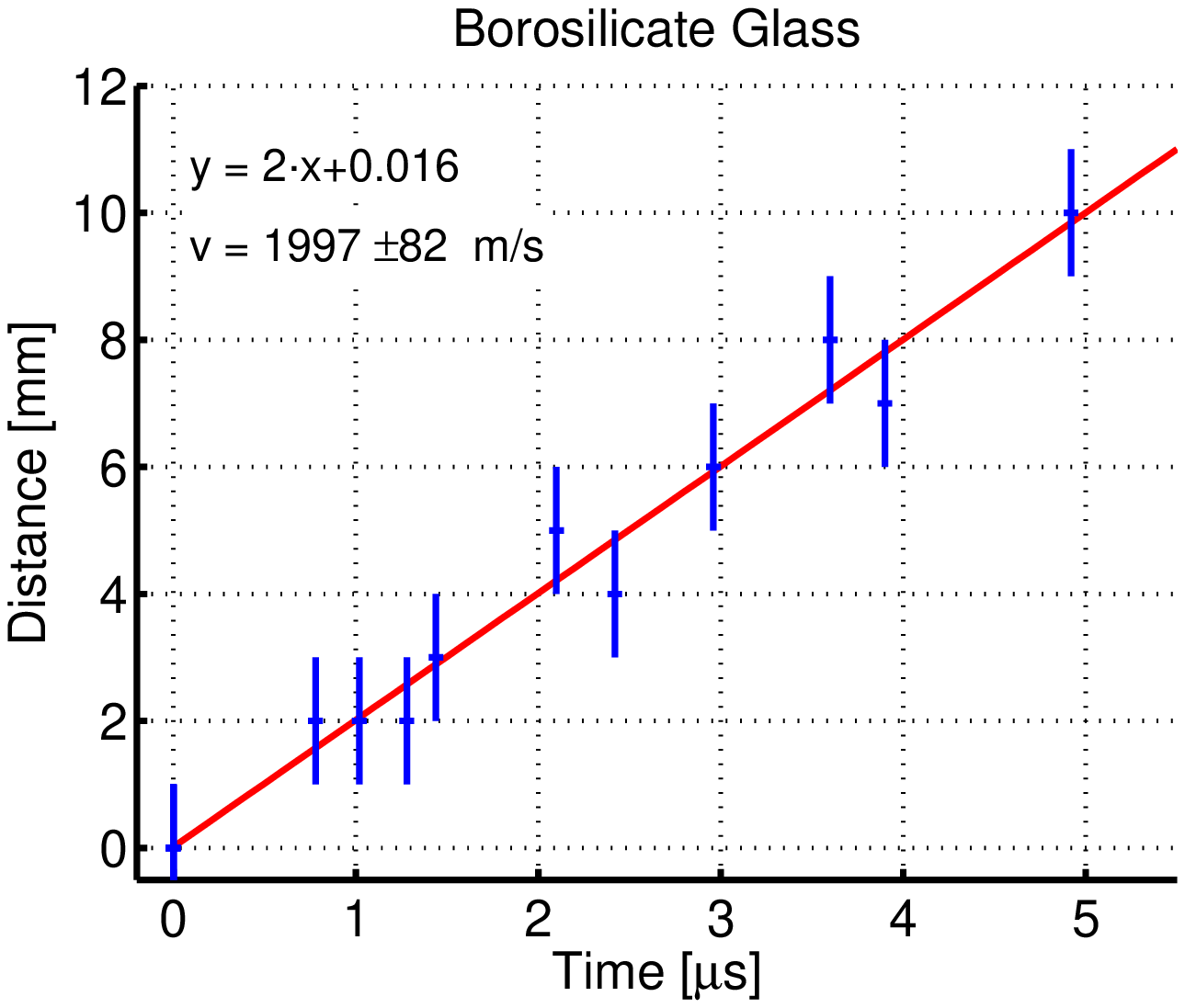}}
		\subfloat[]{
			\label{fig:fig4-b}
			\includegraphics[width=0.33\textwidth]{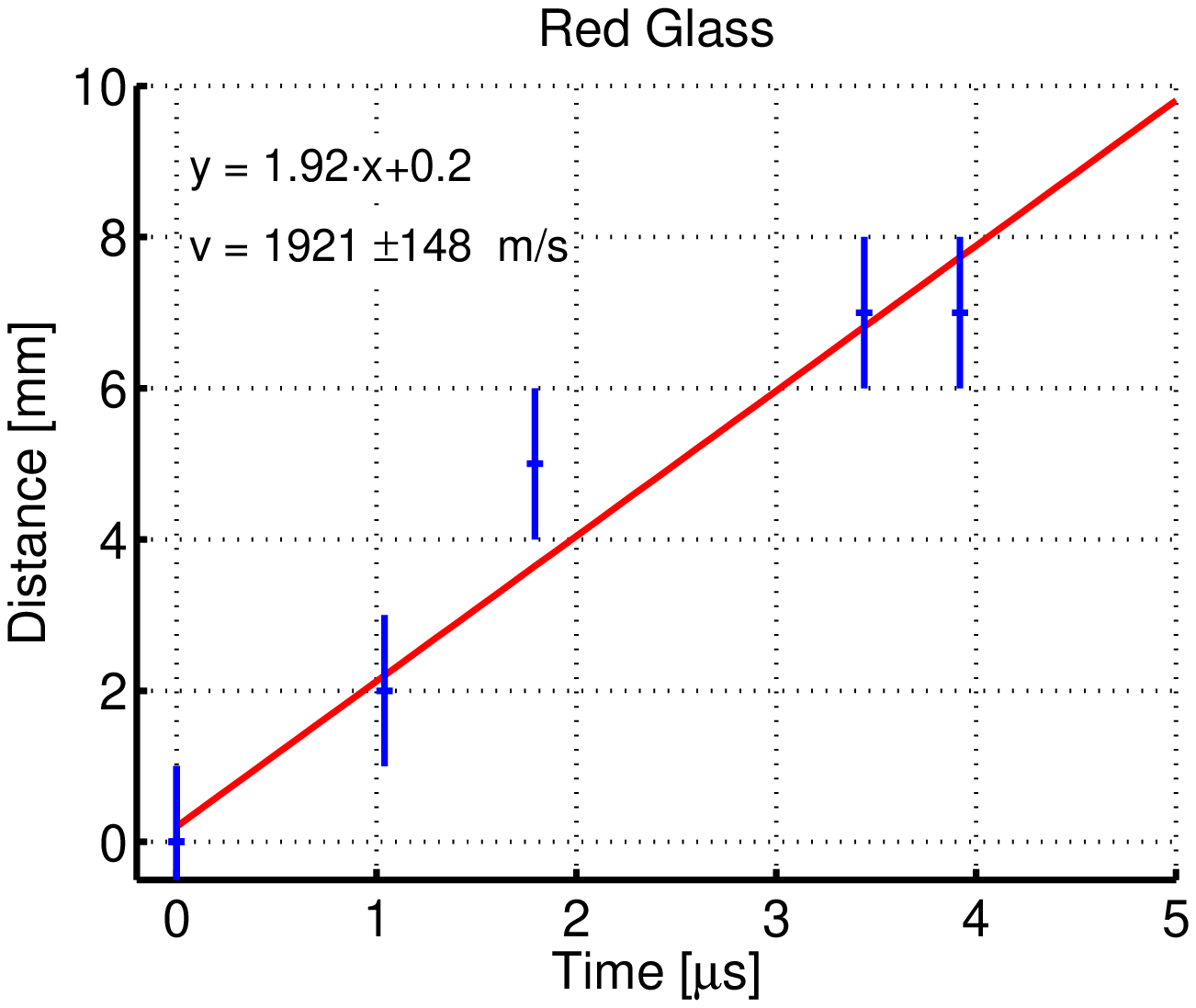}}
		\subfloat[]{
			\label{fig:fig4-c}
			\includegraphics[width=0.33\textwidth]{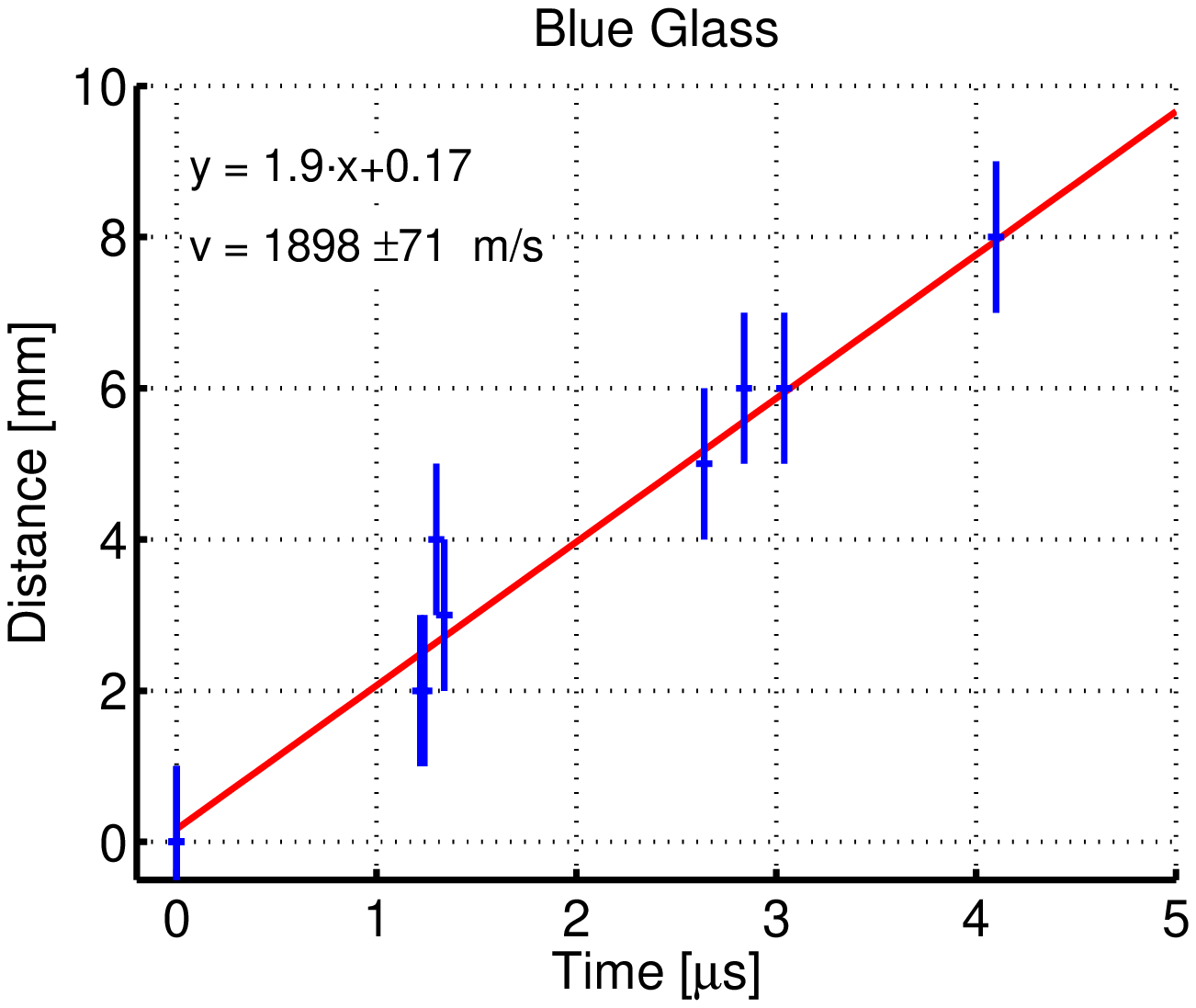}}
		\caption{Disintegration speed of Prince Rupert's drops, estimated from measured oscilloscope traces, for the borosilicate (a), blue (b) and red glass (c). All the samples for this experiment were cooled at T$_{cool} \approx$ 18 $^{\circ}$C.}
		\label{fig:fig4}
	\end{figure}

\begin{figure}[ht]
		\centering
		\subfloat[T$_{cool} \approx$ 18 $^{\circ}$C]{
			\label{fig:fig5-a}
			\includegraphics[width=0.45\columnwidth]{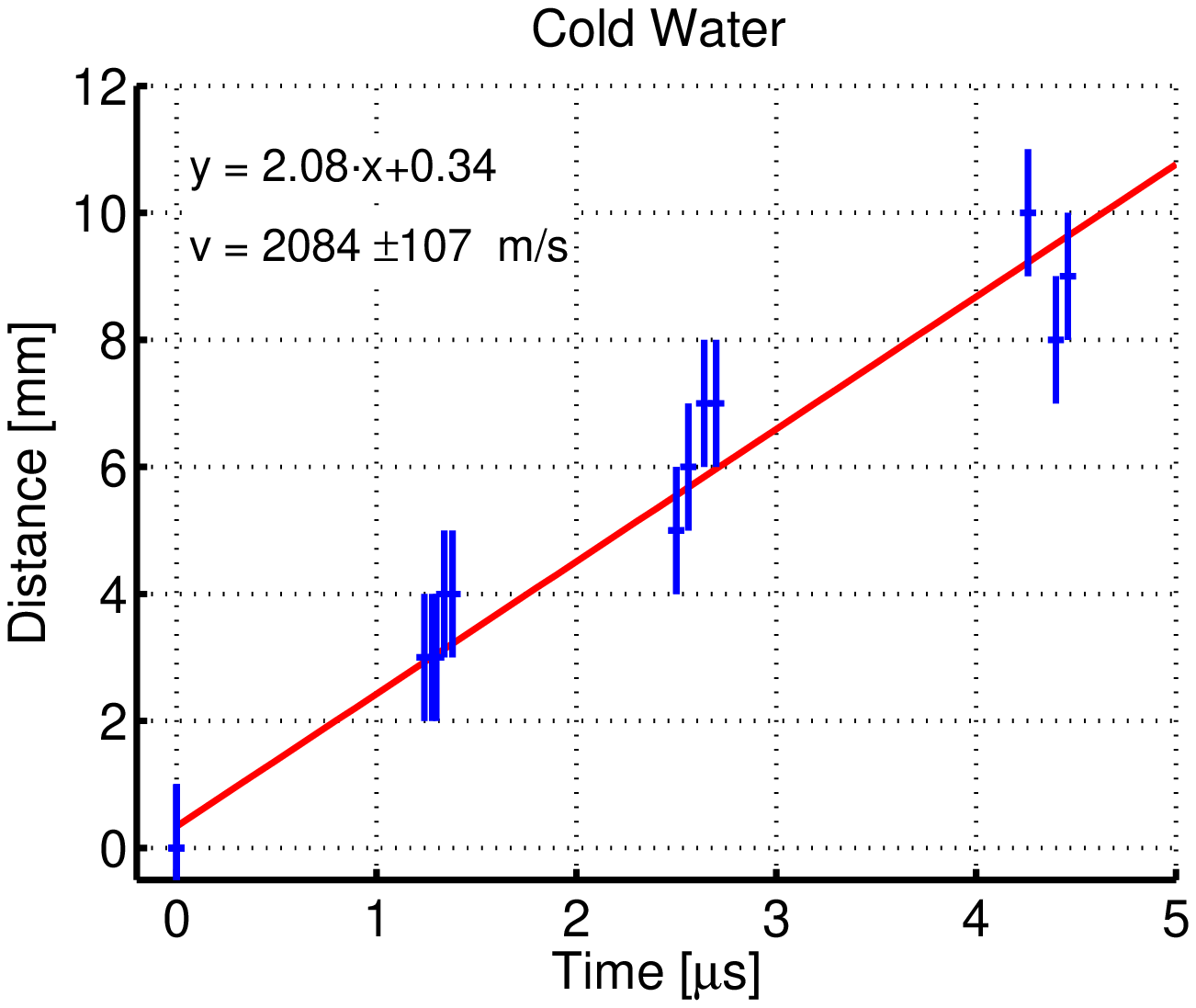}}
		\subfloat[T$_{cool} \approx$ 60 $^{\circ}$C]{
			\label{fig:fig5-b}
			\includegraphics[width=0.45\columnwidth]{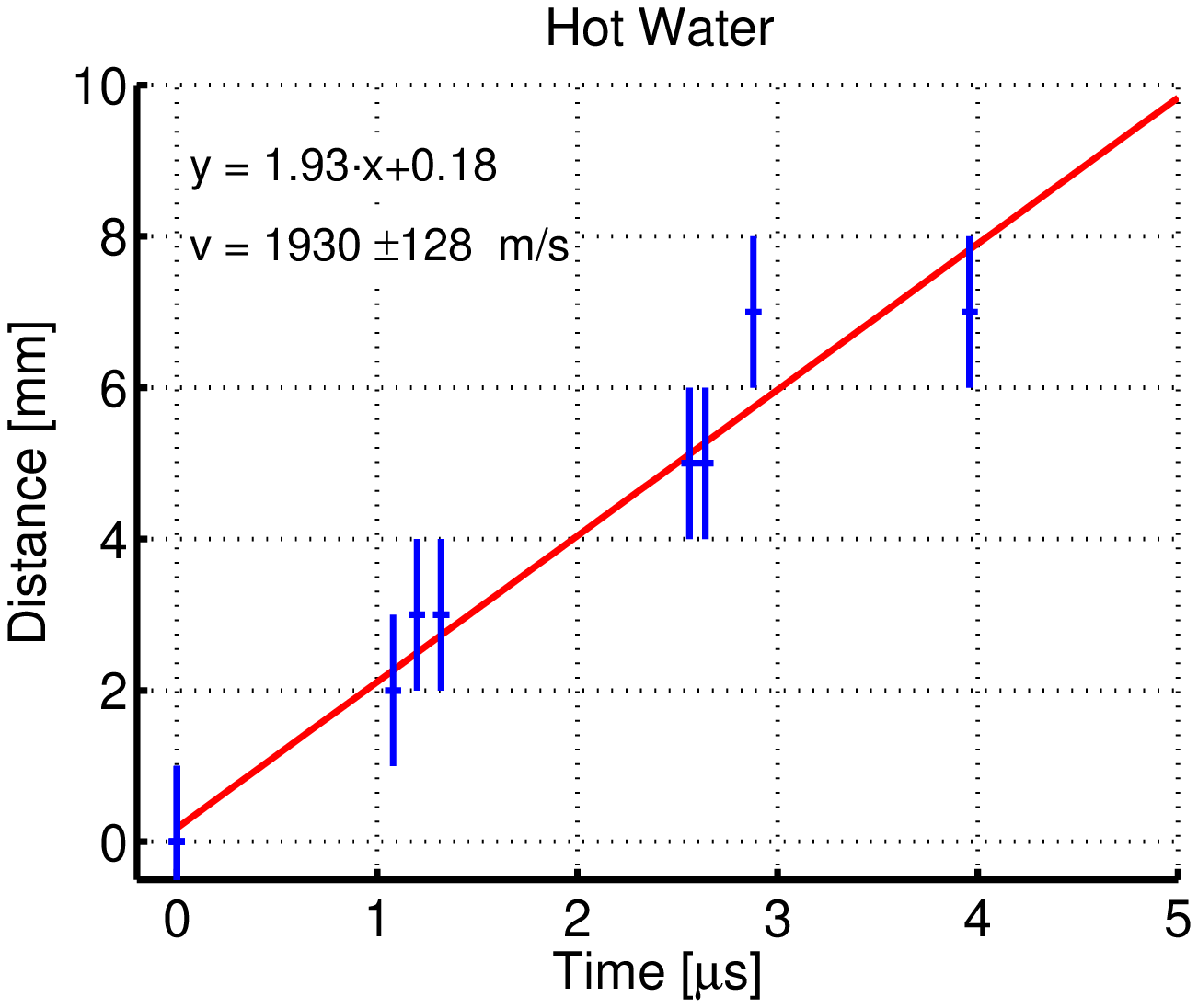}}
		\caption{Disintegration speed of Prince Rupert's drops, fabricated out of soda-lime glass, when cooling the drops in cold (a) and in hot water (b).}
		\label{fig:fig5}
	\end{figure}

	We conducted the same experiment independently varying the type of glass, from which the samples were made, and cooling conditions during the formation of the drop and estimated disintegration speed for each case. The main results are presented in the figure~\ref{fig:fig4}, figure~\ref{fig:fig5} and in the Table.~\ref{table:speeds}. All the measured velocities are around $2000~m/s$, which is in good agreement with $1450-1900~m/s$, measured in \cite{Chandrasekar1994,Chaudhri1998}, and with theoretical estimates, mentioned earlier, resulting in velocities from $1265~m/s$ to $2145~m/s$ for various types of glass.
	
	As one can see from figure~\ref{fig:fig4}, the measured disintegration velocities do not depend much on the material used to fabricate the samples. However, with a better measurement precision we would expect to see lower fracture wave velocity in glasses containing more inclusions of heavy elements, as the sound velocity $c_0 \propto \sqrt{1/\rho}$. This might be the reason of higher disintegration speed in soda-lime glass (figure~\ref{fig:fig5-a}) in comparison to the other three types of glass with heavier inclusions. For the case of different cooling conditions, shown in figure~\ref{fig:fig5}, we observe a higher disintegration speed when cooling the drops in cold water, due to larger stresses in the produced samples.
	
	We tried to make the temperature contrast even larger by cooling the falling drops in liquid nitrogen at $T \approx 77 $ K, but the produced drops were not disintegrating in an explosive manner as regular Prince Rupert's drops do. This probably happens due to fast evaporation of liquid nitrogen around the drop which prevents further heat exchange, so the drop is cooling more steadily than in water, and the resulting stress inside the glass is not enough for explosive destruction.
	
	\begin{table}[ht]
		\caption{Experimentally measured disintegration velocities of Prince Rupert's drops for all types of glass and cooling conditions, used in the experiment.}
		\begin{indented}
			\item[]\begin{tabular}{@{}ll}
				\br
				Type of glass & Estimated speed, km/s \\
				\mr
				borosilicate & 1.997 $\pm$  0.082  \\
				red & 1.921 $\pm$  0.148  \\
				blue & 1.898 $\pm$  0.071  \\
				soda-lime (cold water) & 2.084 $\pm$  0.107  \\
				soda-lime (hot water) & 1.930 $\pm$  0.128  \\
				\br
			\end{tabular}
		\end{indented}
		\label{table:speeds}
	\end{table}

	\section{\label{sec:discuss}Discussion}
	
	As we have shown, such simple setup as in figure~\ref{fig:fig2} allows measuring the disintegration speed of Prince Rupert's drops using the resistance grid technique. However, to achieve maximum precision using this approach one should consider the following remarks.
	
	The biggest influence on the measurement results is made by relations between control resistors. If their nominal values are chosen equal as shown above it leads to a significant decrease of signal amplitude as the fracture wave propagates. When the $n^{th}$ bus connection is broken the corresponding voltage step decreases as $n^{-1}$.
	
	In order to achieve equal per-step voltage drop one should use control resistors with different nominal values. A simple algorithm could be used to estimate maximal number of steps achievable for a specific setup. For this estimation we assume that time distortion of a single step is negligible compared to its length and resistance of the conductive bus is close to zero. 
	
	In this case the only limiting factor is the electric noise that can be seen on the screen of the oscilloscope. Having its amplitude one should define minimal voltage step that can be distinguished from noise. This level corresponds to a single step on the plot which limits the maximal number of observable steps N. Applying Ohm's law to the circuit one can achieve a non-recurrent expression for the value of $n^{th}$ control resistor:
	
	\begin{equation}
		R_n = \frac{N}{(N-n)(N-n+1)}R_0
	\end{equation}

	However, considering small size of a typical object of investigation it would not always be easy to achieve this limit due to both high contacts density and step front distortion.
	
	Second biggest point of consideration is decreasing front distortions. This effect corresponds to the fact that bus connection with the linking wires takes some time to be broken. It depends on the physical size of the wire, width of the conducting bus strip, the thickness of its layer, physical dimensions of a glue drop that links the wire with the bus and on the fact that the disintegration front itself is distorted.
	
	There are several ways to suppress this effect. One should consider making both thickness and width of the conductive bus on the surface of the drop as small as possible. The best way to achieve this might be to deposit the bus as a metal film by evaporation. Another way is using mechanical fixtures instead of glue, proposed in our experiment. But this method requires being careful and keeping stress on the drop's walls several orders of magnitude less than its inner tension. Otherwise it may result in local stresses that may distort the propagation speed of the fracture wave.
	
	Another way to get more data from a single measurement is to track several paths simultaneously. An easy way to do it would be to connect several circuits to a multi-channel oscilloscope, but to encode different paths into various height of the voltage steps using several control sub circuits connected in parallel to one power source. There is one principal limitation of the technique used: it is impossible to measure propagation time of the wavefront between two last contacts (closest to the 'head' of the drop). One should develop the map of contacts bearing this fact in mind.
	
	An important point to take care of is safety during the experiment described, as Prince Rupert's drops are disintegrating into numerous sub-millimeter size shards traveling at very high velocities. In order to use the setup for demonstration and training lab experiments, a transparent shield should be added to prevent small pieces of glass from scattering around. Also it is strongly advisable to use protective goggles and gloves when dealing with Prince Rupert's drops.
	
	\section{\label{sec:concl}Conclusion}
	
	We showed that the resistance grid technique can be used to measure disintegration speed of prince Rupert's drops. Moreover, the obtained results are in good agreement with theoretical estimates and are close to the values, obtained using other measurement methods, such as filming the process with a high-speed camera. We also discussed how to make further improvements to achieve even higher measurement precision.
	
	The proposed method of measuring fast processes could be used not only to measure the disintegration speed of Prince Rupert's drops, but of any other fast disintegration processes such as various explosions. Aside from scientific research this method is ideally suited for demonstration and training experiments in a laboratory setting at universities and high-schools, due to its simplicity, low cost and visual attraction.
	
	\ack
	
	We would like to thank A. Afanasyev for glass supplies, A. Glushkova for visual materials and V. Mislavskiy for helping us with fabrication of Prince Rupert's drops.
	
	\section*{References}
	\bibliography{main}

\begin{thebibliography}{10}

\bibitem{hooke1665}
Robert Hooke.
\newblock {\em Micrographia: or some physiological descriptions of minute
  bodies made by magnifying glasses, with observations and inquiries
  thereupon}.
\newblock Courier Corporation, 2003.

\bibitem{brodsley1986}
Laurel Brodsley, Charles Frank, and John~W Steeds.
\newblock Prince rupert's drops.
\newblock {\em Notes and Records of the Royal Society of London}, pages 1--26,
  1986.

\bibitem{Johnson1992}
W.~Johnson and S.~Chandrasekar.
\newblock {Rupert's glass drops: Residual-stress measurements and calculations
  and hypotheses for explaining disintegrating fracture}.
\newblock {\em Journal of Materials Processing Technology}, 31(3):413--440,
  1992.

\bibitem{Chandrasekar1994}
S.~Chandrasekar and M.~M. Chaudhri.
\newblock {The explosive disintegration of Prince Rupert's drops}.
\newblock {\em Philosophical Magazine Part B}, 70(6):1195--1218, 1994.

\bibitem{Anthony1970}
S.~R. Anthony, J.~P. Chubb, and J.~Congleton.
\newblock {The crack-branching velocity}, 1970.

\bibitem{vanovskiy2014}
Vladimir Vanovskiy.
\newblock International physicists’ tournament—the team competition in
  physics for university students.
\newblock {\em European Journal of Physics}, 35(6):064003, 2014.

\bibitem{griffith1921}
Alan~A Griffith.
\newblock The phenomena of rupture and flow in solids.
\newblock {\em Philosophical transactions of the royal society of London.},
  pages 163--198, 1921.

\bibitem{mott1948}
N~F Mott.
\newblock Brittle fracture in mild steel plates.
\newblock {\em Engineering}, 165:16--18, 1948.

\bibitem{yoffe1951}
Elizabeth~H Yoffe.
\newblock Lxxv. the moving griffith crack.
\newblock {\em The London, Edinburgh, and Dublin Philosophical Magazine and
  Journal of Science}, 42(330):739--750, 1951.

\bibitem{roberts1954}
DK~Roberts and AA~Wells.
\newblock The velocity of brittle fracture.
\newblock {\em Engineering}, 178(4639):820--821, 1954.

\bibitem{steverding1970}
B~Steverding and SH~Lehnigk.
\newblock Response of cracks to impact.
\newblock {\em Journal of Applied Physics}, 41(5):2096--2099, 1970.

\bibitem{bouchbinder2010}
Eran Bouchbinder, Jay Fineberg, and M~Marder.
\newblock Dynamics of simple cracks.
\newblock {\em Annual Review of Condensed Matter Physics}, 1:371--395, 2010.

\bibitem{mosolov1991}
AB~Mosolov.
\newblock Cracks with fractal surfaces.
\newblock In {\em Dokl Akad Nauk SSSR}, volume 319, pages 840--4, 1991.

\bibitem{cherepanov1995}
Genady~P Cherepanov, Alexander~S Balankin, and Vera~S Ivanova.
\newblock Fractal fracture mechanics—a review.
\newblock {\em Engineering Fracture Mechanics}, 51(6):997--1033, 1995.

\bibitem{Yavari2010}
Arash Yavari and Hamed Khezrzadeh.
\newblock {Estimating terminal velocity of rough cracks in the framework of
  discrete fractal fracture mechanics}.
\newblock {\em Engineering Fracture Mechanics}, 77(10):1516--1526, 2010.

\bibitem{galin1966}
LA~Galin and GP~Cherepanov.
\newblock Self-sustaining failure of a stressed brittle body.
\newblock In {\em Soviet Physics Doklady}, volume~11, page 267, 1966.

\bibitem{cherepanov2009}
Genady~P Cherepanov.
\newblock Fracture waves revisited.
\newblock {\em International journal of fracture}, 159(1):81--84, 2009.

\bibitem{bless2010}
Stephan~J Bless.
\newblock Failure waves and their possible roles in determining penetration
  resistance of glass.
\newblock {\em International Journal of Applied Ceramic Technology},
  7(3):400--408, 2010.

\bibitem{Chaudhri1998}
M.~Munawar Chaudhri.
\newblock {Crack bifurcation in disintegrating Prince Rupert's drops}.
\newblock {\em Philosophical Magazine Letters}, 78(2):153--158, 1998.

\bibitem{silverman2012}
MP~Silverman, W~Strange, J~Bower, and L~Ikejimba.
\newblock Fragmentation of explosively metastable glass.
\newblock {\em Physica Scripta}, 85(6):065403, 2012.

\bibitem{herrmann1989}
HJ~Herrmann, J~Kert{\'e}sz, and L~De~Arcangelis.
\newblock Fractal shapes of deterministic cracks.
\newblock {\em EPL (Europhysics Letters)}, 10(2):147, 1989.

\bibitem{herrmann1991}
HJ~Herrmann.
\newblock Patterns and scaling in fracture.
\newblock {\em Physica Scripta}, 1991(T38):13, 1991.

\end{thebibliography}
	
\end{document}